\documentclass[8pt]{article}  \usepackage{times}
\usepackage{graphicx}

\usepackage{color} %--inserted by me

\begin{document}


%\baselineskip24pt


%\maketitle

%\date{}


\twocolumn[{\Huge \textbf{On the action potential as a propagating 
density pulse and the role of anesthetics\\*[0.1cm]}}\\

{\large Thomas Heimburg,$^{\ast}$ and Andrew D. Jackson\\*[0.1cm]
{\normalsize The Niels Bohr Institute, University of Copenhagen,
Blegdamsvej 17, 2100 Copenhagen \O, Denmark\\} {\normalsize
$^{\ast}$corresponding author, theimbu@nbi.dk\\*[0.3cm]} {\normalsize
keywords: Action potential; Hodgkin-Huxley model; solitons; heat
changes\\*[0.3cm]}

{\normalsize The Hodgkin-Huxley model of nerve pulse propagation relies on ion
currents through specific resistors called ion channels. We discuss a
number of classical thermodynamic findings on nerves that are not
contained in this classical theory. Particularly striking is the
finding of reversible heat changes, thickness and phase changes of
the membrane during the action potential. Data on various nerves
rather suggest that a reversible density pulse accompanies the action
potential of nerves. Here, we attempted to explain these phenomena by
propagating solitons that depend on the presence of cooperative phase
transitions in the nerve membrane. These transitions are, however,
strongly influenced by the presence of anesthetics. Therefore, the
thermodynamic theory of nerve pulses suggests a explanation for the
famous Meyer-Overton rule that states that the critical anesthetic
dose is linearly related to the solubility of the drug in the
membranes. \\*[0.5cm] }}]

%------------------------------------------------------------------
\section{Introduction}\label{sec1}
The description of electrical phenomena in nerves is among the first
biological problems studied in physics.  Galvani \cite{Galvani1791}
noticed that the legs of dissected frogs made active movements when
their nerves were connected to a battery.  He called this phenomenon
``animal electricity''.  After learning about these experiments, Volta
\cite{Volta1900} stated that nerve pulses are electrical conduction
phenomena.  Helmholtz \cite{Helmholtz1852} performed the first
measurements of the propagation velocity of nerves and found a value
of about 30 m/s in the nerves from frog muscle.  In the second half of
the 19th century Ostwald \cite{Ostwald1890} and others developed the
theory of osmosis and electrochemistry, and attempts were made to
relate the flux of ions through the nerve membranes to the propagating
action potential \cite{Bernstein1912}.  This finally resulted in the
model by Hodgkin and Huxley \cite{Hodgkin1952} from 1952 that is the
presently accepted model for the nerve pulse.  This model relies on
ionic currents through ion-selective objects (ion channel proteins)
and the membrane capacitor.  In the context of their model, the
conductance of these objects displays rather complex voltage and time
dependences that enter the differential equation via a set of
empirical parameters.  Those parameters are taken from experiment but
do not yet have a satisfying theoretical justification.  Even though
Hodgkin and Huxley \cite{Hodgkin1952} did not originally specify the
ion-conducting objects, it was clear from the line of argument that
these objects were expected to be specific proteins called
ion-channels.  In 1976, Neher and Sakmann using the patch clamp
technique described such channels microscopically \cite{Neher1976}.
Nowadays, many investigators all over the world investigate the
properties of ion channels.  In 1998, MacKinnon and collaborators
crystallized the potassium channel and suggested a pathway for the
potassium through a pore within the protein \cite{Doyle1998}.  Thus,
the Hodgkin-Huxley model seemingly finds support in independent
experiments.  The model by Hodgkin and Huxley is a purely electrical
description based on conductors (ion channels and the cytosol of the
nerve axon) and on a capacitor, which is the lipid membrane.  It does
not contain any thermodynamical variable except the membrane
potential.  Entropy, temperature, pressure and volume do not play a
role.  There is, however, strong evidence that phenomena during the
action potential are not purely electrical.  It has been observed by a
number of investigators that the dimensions of the nerve change in
phase with voltage changes and that the nerve exerts a force normal to
the membrane surface
\cite{Iwasa1980a,Iwasa1980b,Tasaki1990,Tasaki1982b,Tasaki1980}.
Further, during the action potential lipid membrane markers change
their fluorescence intensity and their anisotropy
\cite{Tasaki1969a,Tasaki1968}.  Most striking, however, is the finding
that there are reversible changes in temperature and heat during the
action potential
\cite{Abbott1958,Howarth1968,Ritchie1985,Tasaki1989,Tasaki1992a}.
While the Hodgkin-Huxley model \cite{Hodgkin1952} contains resistors
that should generate heat during the flow of ions, the reversible
release and re-absorption of heat does not find a satisfactory
explanation within this model \cite{Hodgkin1964}.  Recently, Heimburg
and Jackson \cite{Heimburg2005c,Lautrup2005} proposed that the action
potential is rather a propagating density pulse (soliton), and
therefore an electromechanical rather than a purely electrical
phenomenon.  This corresponds to a localized piezoelectric sound pulse
within the nerve membrane.  Such a model is able to explain most of
the thermodynamical findings on nerves and results in the correct
propagation velocity of about 100 m/s for a myelinated nerve.
Interestingly, Hodgkin and Huxley themselves proposed the possibility
that the nerve pulse is a propagating mechanical wave 
\cite{Hodgkin1945}.
%vvvvvvvvvvvvvvvvvvvvvvvvvvvvvvvvvvv
\begin{figure*}[t!]
    \begin{center}
	\includegraphics[width=12cm]{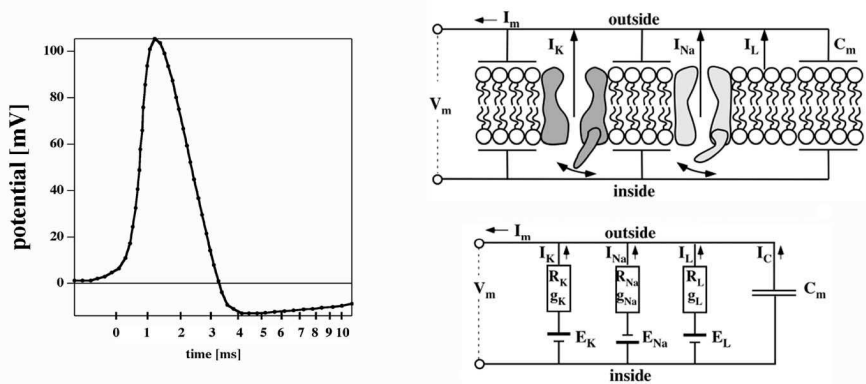}
	\parbox[c]{14cm}{ \caption{\textit{Left: Action potential
	adapted from the original paper of Hodgkin and Huxley
	\cite{Hodgkin1952}.  Right top: Electrical currents in the
	Hodgkin-Huxley model through ion channels.  Right bottom:
	Equivalent circuit picture replacing ion channels by resistors
	and the membrane as a capacitor.}
	\label{fig:Figure1}}}
    \end{center}
\end{figure*}
%^^^^^^^^^^^^^^^^^^^^^^^^^^^^^^^^^^^
Anesthesia is a phenomenon that seems to be closely related to the
action of nerves.  Since the standard model of nerve action is based
on the action of ion channels, most research has been dedicated to
investigating the influence of anesthetics on such proteins.  However,
an old finding by Meyer \cite{Meyer1899} and Overton
\cite{Overton1901,Overton1991} states that the action of anesthetics
is linearly related to their solubility in membranes.  This includes
the noble gas Xenon.  Although some ion channels are influenced by
some anesthetics, there is no quantitative correlation with the
well-documented Meyer-Overton rule \cite{Campagna2003}.

In this paper we briefly discuss some of the historical findings on
nerves, including the Hodgkin-Huxley model and thermodynamic data on
nerves. It is shown that the Hodgkin-Huxley theory does not describe
the thermodynamics of the nerve pulse correctly. Instead, the
propagation of a density pulse is shown to explain in a quantitative
manner many features of the nerve pulse, including density,
fluorescence anisotropy and heat changes. Finally, we show that such
a description leads to a satisfactory quantitative explanation of
general anesthesia.

%------------------------------------------------------------------
\section{The Hodgkin-Huxley model}\label{sec2}
In the Hodgkin-Huxley model \cite{Hodgkin1952} the propagation of a
voltage pulse is the consequence of ion currents through the membrane
and along the nerve axon.  The electrochemical potential (Nernst
potential) across the nerve membrane balances the ion concentration
differences on both sides of the nerve axon.  The transient opening of
voltage-dependent ion channels leads to a related transient voltage
change that can propagate.  Most of the data on which the
Hodgkin-Huxley model is based originate from voltage-clamp experiments
on giant squid axons where the trans-membrane voltage is kept constant
along the whole length of the axon.

The relation for the ion current through the membrane under voltage
clamp conditions is based on an equivalent circuit picture that is
schematically shown in Fig. 1. Describing ion channels by resistors
and the membrane as a capacitor, one obtains
\begin{equation}
    I_{m}=C_{m}\frac{dU}{dt}+g_{K}(U-E_{K})+g_{Na}(U-E_{Na})+g_{L}(U-E_{L})
    \label{2.1}
\end{equation}
where $I_{m}$ is the current through the membrane, and $C_{m}$ is the
capacitance of the membrane (typically on the order of $1mF/cm^2$). The
$E_{K}$, $E_{Na}$ and $E_{L}$ are resting potentials that depend on ion
concentrations. The $g_{K}$ and $g_{Na}$ are the conductances of K-channels and
Na-channels, and $g_{L}$ describes the leakage currents. The conductances
are not constants but rather complicated functions of time and
voltage, $g_{K}=g_{K}(V,t)$ and $g_{Na}=g_{Na}(V,t)$, that have been empirically
fitted by Hodgkin and Huxley \cite{Hodgkin1952} using many ad hoc parameters.
Therefore, the seemingly simple eq. (\ref{2.1}) is in fact very complicated,
and all the mysteries of the observed phenomena are hidden in the
functional dependences of the conductances on time and voltage.
The trans-membrane current in eq. (1) is given as the sum of a
capacitive current and an Ohmic current. The capacitive current is
given by
\begin{equation}
    I_{C}=\frac{d}{dt}\left(C_{m}\cdot U\right)=
    C_{m}\frac{dU}{dt}+U\frac{dC_{m}}{dt}
    \label{2.2}
\end{equation}
A closer look at the right hand side of eq. (\ref{2.1}) indicates that the
capacitive current used by Hodgkin and Huxley consists only of the
$C_{m}\cdot dU/dt$ term and that the capacitance $C_{m}$ was assumed to be constant.
Therefore the $U\cdot dC_{m}/dt$ term has been neglected. This is probably not
correct since we will show in the next section that the thickness of
nerves changes during the pulse. Note in particular that the function
dCm/dt carries the same units as the conductances, $g_{i}$. For this
reason it may not always be trivial to distinguish currents through
resistors and capacitive currents in an experiment during a
propagating pulse \cite{Kaufmann1989d,Kaufmann1989e}.
%vvvvvvvvvvvvvvvvvvvvvvvvvvvvvvvvvvv
\begin{figure*}[htb!]
    \begin{center}
	\includegraphics[width=12cm]{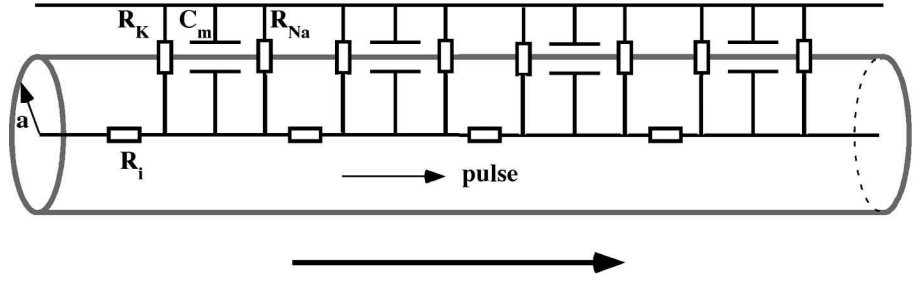}
	\parbox[c]{14cm}{ \caption{\textit{Equivalent circuit picture
	of a propagating voltage pulse.  Currents flow along the nerve
	axon and across the axonal membrane through resistors and
	should produce net heat dissipation.}
	\label{fig:Figure2}}}
    \end{center}
\end{figure*}
%^^^^^^^^^^^^^^^^^^^^^^^^^^^^^^^^^^^
To arrive at a wave equation for the nerve axon, Hodgkin and Huxley
assumed that the total current is the sum of the trans-membrane
current and the current along the axon. A further ad-hoc assumption
is that a propagating solution exists that fulfills a wave equation.
Hodgkin and Huxley6 arrived at the following differential equation
for the propagating nerve pulse:
\begin{equation}
    \frac{a}{2R_{i}}\frac{\partial^2 U}{\partial x^2}=
    C_{m}\frac{dU}{dt}+g_{K}(U-E_{K})+g_{Na}(U-E_{Na})
    \label{2.3}
\end{equation}
where $a$ is the radius of the axon and $R_{i}$ is the resistance of the
cytosol within the nerve. This equation introduces a dependence of
the pulse propagation on the nerve radius. The elements of the
propagating pulse are summarized in Fig. 2 that shows the equivalent
circuits as an in-line arrangement of many local equivalent circuits
as shown in Fig. 1. Due to the voltage and time dependence of the
conductances in eq. (\ref{2.3}) this differential equation can only be
solved numerically. Hodgkin and Huxley found a convincing agreement
between the calculated and the observed pulse shape for the squid
axon that only contains K- and Na-currents.\\

One immediate implication of the Hodgkin-Huxley model is that ion
currents through the nerves should produce heat. Electrical currents
through resistors generate heat, independent of the direction of the
ion flux. The heat production in such an experiment therefore should
always be positive if the Hodgkin-Huxley model is taken seriously and
the analogy of ion currents through protein pores and Ohmic currents
is assumed to be correct. The heat dissipation should be related to
the power of a circuit through the resistor, i.e. $dQ/dt=P=U\cdot I=
g^{-1}I^2>0$ for each of the conducting objects in all phases of the
action potential. In the next section we will show that this is not
in agreement with the experiment.

%------------------------------------------------------------------
\section{Thermodynamics of nerve pulses}\label{sec3}
The Hodgkin-Huxley model \cite{Hodgkin1952} is a purely electrical
theory.  It is based on equivalent circuits and makes use of
capacitance, resistors and ionic currents.  It is not a thermodynamic
theory.  It does not explicitly contain temperature and heat or other
thermodynamic variables such as pressure, volume and the chemical
potentials of molecules dissolved in the membrane (e.g. anesthetics).
However, there are many reports in the literature indicating that, in
addition to the electrical response of nerves, other variables also
change, for example the thickness, the enthalpy and heat content of
the nerve.  In the following we briefly discuss some of these data.

%'''''''''''''''''''''''''''''''''''''''''''''''''''''
\subsection{Thickness and forces}\label{sec3.1}

%vvvvvvvvvvvvvvvvvvvvvvvvvvvvvvvvvvv
\begin{figure*}[htb!]
    \begin{center}
	\includegraphics[width=12cm]{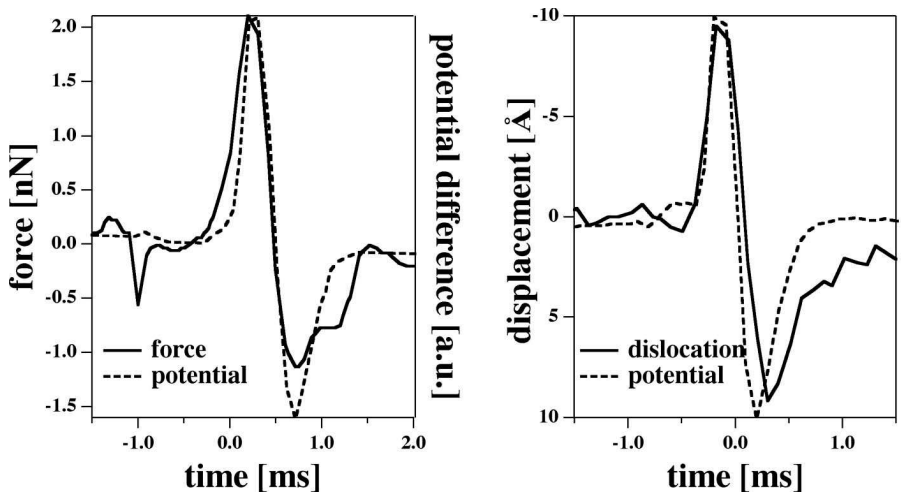}
	\parbox[c]{14cm}{ \caption{\textit{Mechanical changes during
	the action potential.  Left: Force on a piston during the
	action potential in a squid axon.  The solid line represents
	the voltage changes, the dotted curve the force.  Right:
	During the nerve pulse in a squid axon the thickness of the
	nerve changes proportional to the voltage.  Data on squid
	axons adapted from ref. \cite{Iwasa1980a}.}
	\label{fig:Figure3}}}
    \end{center}
\end{figure*}
%^^^^^^^^^^^^^^^^^^^^^^^^^^^^^^^^^^^
I. Tasaki and collaborators have published several studies on the
mechanical and thermodynamic properties of various nerves
\cite{Iwasa1980a,Iwasa1980b,Tasaki1990,Tasaki1982b,Tasaki1980,Tasaki1969a,Tasaki1989,Tasaki1992a,Kobatake1971}.
For all nerves investigated, they found that the action potential
(i.e. the voltage pulse) is accompanied by changes in the dimensions
of the nerve.  In Fig.3 (right) it is shown that the voltage pulse of
a squid axon is exactly proportional to the change of its thickness
\cite{Iwasa1980a,Iwasa1980b}.  In the example this thickness change is
about 1 nm.  Further, the same authors showed that during this pulse a
considerable force acts on a piston that was brought into contact with
the nerve surface (Fig.3, left).  The force on that piston (0.01
cm$^2$ cross section) was shown to be about 2 nN at the voltage peak
maximum.

%'''''''''''''''''''''''''''''''''''''''''''''''''''''
\subsection{Fluorescence changes, optical changes and alterations in lipid
state}\label{sec3.2}

During the action potential not only thickness and pressure on a
piston change but also the state of the membrane as measured by the
fluorescence changes of lipid dyes.  Tasaki and coworkers
\cite{Tasaki1969a,Tasaki1968} found that in various nerves under the
influence of the action potential the fluorescence intensity change is
proportional to the voltage pulse (see Figure 4).
 
%vvvvvvvvvvvvvvvvvvvvvvvvvvvvvvvvvvv
\begin{figure*}[htb!]
    \begin{center}
	\includegraphics[width=12cm]{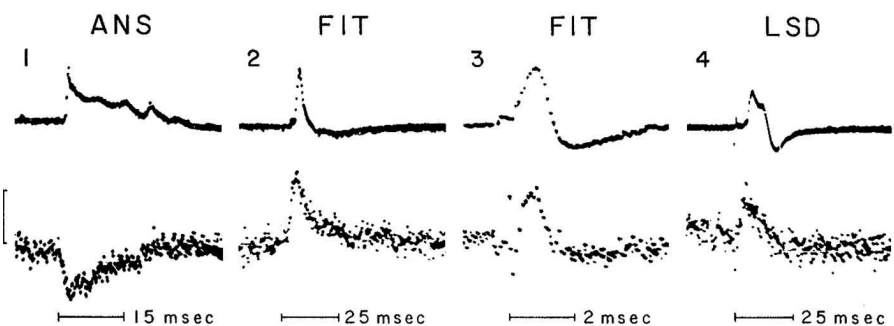}
	\parbox[c]{14cm}{ \caption{\textit{Voltage changes (top
	traces) and fluorescence changes (bottom traces) for 4
	different fluorescence markers and nerve preparations.  They
	are exactly in phase.  1.  Squid giant axon and
	8-anilinonaphtalene-1-sulfonate (ANS).  2.  Crab leg nerve
	with fluorescein isothiocyanate (FIT).  3.  Squid axon with
	FIT. 4.  Crab leg nerve with lysergic acid diethylamide (LSD).
	From ref.  \cite{Tasaki1969a} with permission.  The data were
	taken as a proof for changes of the viscosity within the
	membrane during the action potential.}
	\label{fig:Figure4}}}
    \end{center}
\end{figure*}
%^^^^^^^^^^^^^^^^^^^^^^^^^^^^^^^^^^^
In the same paper they showed that the fluorescence ani\-sotropy of
these markers also changes (data not shown). The fluorescence
anisotropy is a measure of the rotational mobility of the
fluorescence markers. A lower anisotropy indicates faster movement,
whereas a high anisotropy indicates slower movement. Since the
fluorescence anisotropy changed during the voltage pulse, Tasaki and
collaborators \cite{Tasaki1969a} concluded that the viscosity of the membrane changes
during the nerve pulse. Note that they published this paper prior to
the Ôfluid mosaic modelÕ by Singer and Nicholson \cite{Singer1972} from 1972 that
established the present view of the biological membrane. The concept
of phase transitions in lipid membranes was not established. One
should conclude from the fluorescence data that significant changes
in the order of the lipid membrane take place. The evidence for phase
transitions during nerve pulses has been discussed in more detail by
Kinnunen and Virtanen \cite{Kinnunen1986} and Tasaki and coworkers 
\cite{Kobatake1971,Tasaki1999}.
In this context it should be noted that also changes in light
scattering and turbidity accompany the action potential that clearly
cannot be related to membrane voltages \cite{Tasaki1969a,Cohen1972}.

%'''''''''''''''''''''''''''''''''''''''''''''''''''''
\subsection{ Reversible heat changes and their meaning}\label{sec3.3}

The most striking thermodynamic findings in nerves during the action
potential are reversible temperature changes and corresponding changes
in the heat released during the nerve pulse.  The first to carefully
describe the heat changes was A. V. Hill who published a series of
papers in the 1920Õs and 1930Õs.  Abbott et al.  \cite{Abbott1958}
showed that the heat release during the first phase of the action
potential is nearly exactly compensated by a heat uptake in the second
phase of the action potential.  This effect was found in
non-myelinated \cite{Abbott1958,Howarth1968,Ritchie1985,Tasaki1989}
and in myelinated \cite{Abbott1958,Tasaki1992a} nerves.  Interestingly,
Hill and collaborators found that the reversible heat release in
myelinated nerves originates from the complete nerve and not only from
the nodes of Ranvier \cite{Abbott1958}.  They found it most likely that the complete
membranes of the myelinated nerves contribute to the heat release and
that one should therefore consider an active role of the myelin sheet
to the nervous impulse.  Saltatory conduction that is the textbook
picture for pulse propagation in myelinated nerves, in contrast,
attributes a special role to the nodes of Ranvier.  Other authors
reproduced the findings on reversible heat release, e.g. Howarth et
al. \cite{Howarth1968}, Ritchie \& Keynes \cite{Ritchie1985} or 
Tasaki and coworkers \cite{Tasaki1989,Tasaki1992a,Tasaki1999}.  It has to
be acknowledged that these experiments are difficult and the observed
temperature changes are small (of order 100mK).\\
%vvvvvvvvvvvvvvvvvvvvvvvvvvvvvvvvvvv
\begin{figure}[htb!]
    \begin{center}
	\includegraphics[width=8cm]{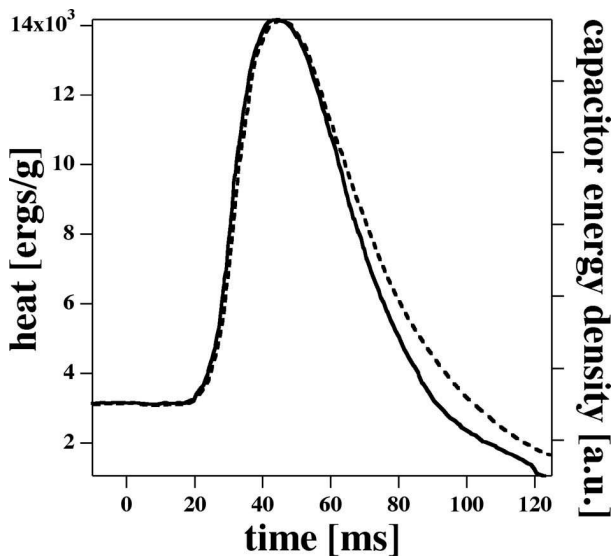}
	\parbox[c]{8cm}{ \caption{\textit{Reversible heat change
	during the action potential.  Left: The square of the voltage
	(the energy of charging a capacitor) is proportional to the
	heat of the nerve pulse.  The heat, however, is much larger
	than the capacitor energy.  The heat during the nerve pulse
	returns to the baseline indicating that the nerve pulse is
	adiabatic (does not generate net heat after completion of the
	action potential).  Data on garfish olfactory nerve adapted
	from ref.  \cite{Ritchie1985}.}
	\label{fig:Figure5}}}
    \end{center}
\end{figure}
%^^^^^^^^^^^^^^^^^^^^^^^^^^^^^^^^^^^
One important result demonstrated in Fig.  5 shows the integrated heat
release during the action potential and the square of the voltage
changes related to the free energy of the membrane capacitor
\cite{Ritchie1985}.  These two functions were found to be
qualitatively nearly identical.  However, the heat reversibly released
during the action potential was several times larger than the energy
of the capacitor so that it can be excluded that the reversible heat
release is explained by the charging of the membrane capacitor.  This
is the only semi-reversible element in the Hodgkin-Huxley model
\cite{Hodgkin1952}.  Further, the heat after the whole pulse returns
to the baseline in phase with voltage changes.  Thus, after the nerve
pulse no net heat was dissipated within experimental error.  Control
experiments indicate that heat is not lost by thermal conduction into
the environment but is rather reabsorbed by the nerve in the second
phase of the action potential.

The reversible heat release is a remarkable and very meaningful
finding. It suggests that the physical processes underlying the nerve
impulse are reversible processes. The Hodgkin-Huxley model, however,
is based on irreversible processes, in particular on the exchange of
potassium and sodium ions along ion gradients. The model does not
contain any true reversible processes. Even if the membrane capacitor
was reversibly char\-ged, this would not result in a reversible heat
change unless the flux of the ions was also reversible, which is not
the case within the framework of the model. Taking the equivalent
circuit picture seriously, the flux of charges through a resistor
should rather result in a heat release independent on the direction
of the flux of the ions. The flux of potassium and of sodium should
both dissipate heat. This is obviously not in agreement with the
thermodynamic results obtained from real nerves. The finding of
changes in lipid state and in thickness also does not find a
satisfactory explanation within the Hodgkin-Huxley model.

%------------------------------------------------------------------
\section{Propagating density pulses}\label{sec4}
In the following we show that the thermodynamic findings described
above find an explanation if one assumes that the action potential
consists of a propagating density pulses. Heimburg and Jackson 
\cite{Heimburg2005c}
showed that one could obtain stable propagating density pulses in
cylindrical lipid membranes provided that the membrane exists in a
physical state slightly above a melting transition. In the following
we outline the underlying basis of this model.

%'''''''''''''''''''''''''''''''''''''''''''''''''''''
\subsection{Melting transitions in biological membrane}\label{sec4.1}
Many biological membranes display melting transitions slightly below
body temperature. In Fig. 6 the melting transition of native \textit{E.coli}
membranes (including all their proteins) are shown. One finds a
pronounced lipid-melting peak slightly below body temperature that is
affected by growth temperature of the bacteria, by hydrostatic
pressure and pH \cite{Heimburg2006c}. Further, one finds several protein unfolding peaks
slightly above body temperature. It is a remarkable fact in itself
that Nature chooses living systems to exist so close to the
cooperative transitions of their molecules, including membranes,
proteins and DNA. The underlying theme of this paper is that this is
of major biological relevance. \\
%vvvvvvvvvvvvvvvvvvvvvvvvvvvvvvvvvvv
\begin{figure*}[htb!]
    \begin{center}
	\includegraphics[width=12cm]{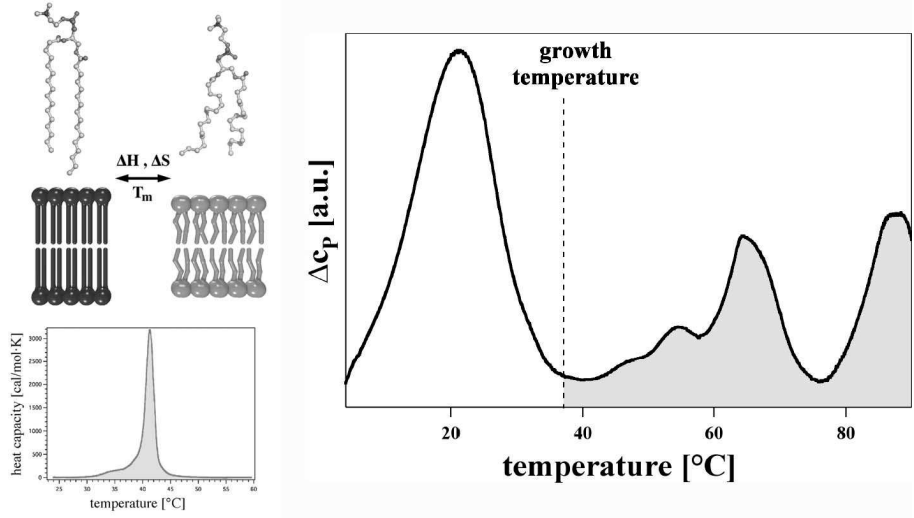}
	\parbox[c]{14cm}{ \caption{\textit{Left: Schematic picture of
	the melting process in lipid membranes and the associated
	change in the specific heat capacity.  Right: Melting profile
	of the membranes of E.coli grown at 37¡C (adapted from ref.
	22).  The growth temperature is indicated as dashed line.  The
	peaks below growth temperature belongs to the melting of lipid
	membranes, the peaks shaded in grey above the growth
	temperature are attributed to protein unfolding.}
	\label{fig:Figure6}}}
    \end{center}
\end{figure*}
%^^^^^^^^^^^^^^^^^^^^^^^^^^^^^^^^^^^
The melting transitions of such membranes display a melting
temperature, Tm, a melting enthalpy, $\Delta H$, and a melting
entropy, $\Delta S$, given by $\Delta S=\Delta H/T_{m}$.  Further,
volume and area of the membrane change during the melting process.
For the model lipid DPPC (dipalmitoyl phosphatidylcholine) that is the
major lipid component of lung surfactant one finds: $T_{m}=314.2$ K,
$\Delta H=35$ kJ/mol, $\Delta S=111.4$ J/mol$\cdot$K, $\Delta
V/V=0.04$ and $\Delta A/A=0.246$.  These values give the order of
magnitude but vary between different lipid species.

%'''''''''''''''''''''''''''''''''''''''''''''''''''''
\subsection{The relation between heat capacity and 
compressibility}\label{sec4.2}
The enthalpy, specific volume and specific area changes in a lipid
melting transition can be written as
\begin{eqnarray}
    H(T) & = & H_{0}(T)+\Delta H(T)\nonumber\\
    V(T) & = & V_{0}(T)+\Delta V(T) \\
    A(T) & = & A_{0}(T)+\Delta A(T)\nonumber
    \label{4.1}
\end{eqnarray}
$H_{0}(T)$ is the temperature-dependent enthalpy of the pure gel phase
and the function $\Delta H(T)$ is the excess enthalpy of the
transition.  Similarly, $V_{0}(T)$ and $A_{0}(T)$ are the
temperature-dependent specific volume and area of the gel phase.
$\Delta V(T)$ and $\Delta A(T)$ are the excess volume and area changes
associated with the melting transition.  It has been found
experimentally that the volume and area changes in the chain melting
transition are proportional to the changes in enthalpy 
\cite{Heimburg1998,Ebel2001}.  
\begin{eqnarray}
    \Delta V(T) & = & \gamma_{V}\cdot \Delta H(T)\nonumber\\
    \Delta A(T) & = & \gamma_{A}\cdot \Delta H(T)
    \label{4.2}
\end{eqnarray}
where the constants $\gamma_{V}=7.8\cdot 10^{-10}\,m^2/N$ and 
$\gamma_{A}=0.89\,m/N$ are approximately
the same for various artificial lipids and for biological membranes.
Using the fluctuation dissipation theorem it is easy to show that
excess heat capacity changes within the lipid melting transition is
proportional to the excess isothermal volume and area compressibility:
\begin{eqnarray}
    \kappa_{T}^V(T)& = & 
    \kappa_{T,0}^V(T)+\frac{\gamma_{V}^2T}{V}\Delta c_{P}\nonumber\\
    \kappa_{T}^A(T)& = & 
	\kappa_{T,0}^A(T)+\frac{\gamma_{A}^2T}{A}\Delta c_{P}
    \label{4.3}
\end{eqnarray}
The heat capacity can easily be measured in calorimetry.  The
functions $\kappa_{T,0}^V$ and  $\kappa_{T,0}^A$ are the temperature dependent
compressibilities of the pure phases that have to be taken from
literature.  One can see that both volume and area compressibilities
assume maxima at the temperature where the heat capacity is maximum.
The adiabatic compressibilities relevant for sound propagation can be
determined when the isothermal compressibilities are known.  They
assume the form \cite{Heimburg1998} 
\begin{eqnarray}
    \kappa_{S}^V(T)& = & 
    \kappa_{T}^V(T)-\frac{T}{V\cdot c_{P}}\left(\frac{dV}{dT}\right)_{P}^2\nonumber\\
    \kappa_{S}^A(T)& = & 
	\kappa_{T}^A(T)-\frac{T}{A\cdot c_{P}}\left(\frac{dA}{dT}\right)_{\Pi}^2
    \label{4.4}
\end{eqnarray}
where the heat capacity $c_{P}$ is that of the
membrane plus the aqueous environment that transiently absorbs heat
from the membrane upon compression.  If the compression is very slow,
$c_{P}$ will be very large and therefore in the limit of very slow
compression $\kappa_{S}^V\approx \kappa_{T}^V$ and 
$\kappa_{S}^A\approx \kappa_{T}^A$. It has been found experimentally that
the adiabatic compressibility obtained for periodic perturbations with
a frequency $\omega=5\,$MHz can be determined accurately if the heat capacity
is assumed be the total heat capacity of the lipid membrane alone.  It
is obviously smaller than the isothermal compressibility.  Therefore,
one has to conclude that the adiabatic compressibility is in general
frequency dependent and, thus, dispersion is present.  The frequency
dependence of relaxation phenomena in the lipid melting transition has
also been documented in experiments \cite{Mitaku1982} and justified 
theoretically \cite{Halstenberg2003}.
It is also obvious from eqs.  (\ref{4.1}) Ð (\ref{4.3}) that the compressibility
is a nonlinear function of the membrane density \cite{Heimburg2005c}.  If the adiabatic
compressibility is known one can calculated the sound velocity, e.g.
for the lateral sound velocity within the membrane plane 
\begin{equation}
    c=\sqrt{\frac{1}{\kappa_{S}^A\rho^A}}
    \label{4.5}
\end{equation}
The lateral area density of the membrane and the enthalpy are related.
Therefore the adiabatic compressibility is a function of the area
density of the membrane, and it follows that the sound velocity is a
nonlinear function of the density that, close to the lipid melting
transition, can be expanded into a power series such that 
\begin{equation}
    c^2=c_{0}^2+p(\Delta \rho^A)+q(\Delta \rho^A)^2+\ldots
    \label{4.6}
\end{equation}
where $c_{0}$ is the sound velocity in the fluid phase of the
membrane.  Here, $p$ and $q$ are parameters to be determined from the
known dependence of the sound velocity on the density.  For
unilamellar DPPC membranes slightly above the transition one finds
experimentally that $c_{0}=176.6$ m/s (the lateral sound velocity in
the fluid phase at low frequencies), $p=-16.6\,c_{0}^2/\rho_{0}^A$ and
$q=79.5\,c_{0}^2/(\rho_{0}^A)^2$ (for details see ref.
\cite{Heimburg2005c}).  Here, $\rho_{0}^A= 4.035\cdot 10^{-3}\,g/m^2$ is
the lateral area density in the fluid phase of the membrane slightly
above the melting point.  Similar values were found for lung
surfactant and native \textit{E.coli} membranes.
 
%'''''''''''''''''''''''''''''''''''''''''''''''''''''
\subsection{Propagating solitons}\label{sec4.3}
We now consider the propagation of a density pulse in a cylindrical
membrane along the axis, x. The hydrodynamic equation for the
propagation of such a density pulse in the presence of
dispersion \cite{Heimburg2005c,Lautrup2005} is given by
\begin{equation}
    \frac{\partial^2}{\partial 
    t^2}\Delta\rho^A=\frac{\partial}{\partial 
    x}\left[c^2\frac{\partial}{\partial x}\Delta 
    \rho^A\right]-h\frac{\partial^4}{\partial x^4}\Delta\rho^A
    \label{4.7}
\end{equation}
describing the changes of the lateral membrane density as a function
of time and space.  The second term is chosen ad hoc to mimic the
frequency dependence of the sound velocity in a linear way using a
parameter h (for details see ref.  \cite{Heimburg2005c}).  This
parameter is the only one that has not yet been determined by
experiment.  We will see below that the only role of the parameter h
is to set the linear scale of the propagating pulse.
%vvvvvvvvvvvvvvvvvvvvvvvvvvvvvvvvvvv
\begin{figure}[htb!]
    \begin{center}
	\includegraphics[width=8cm]{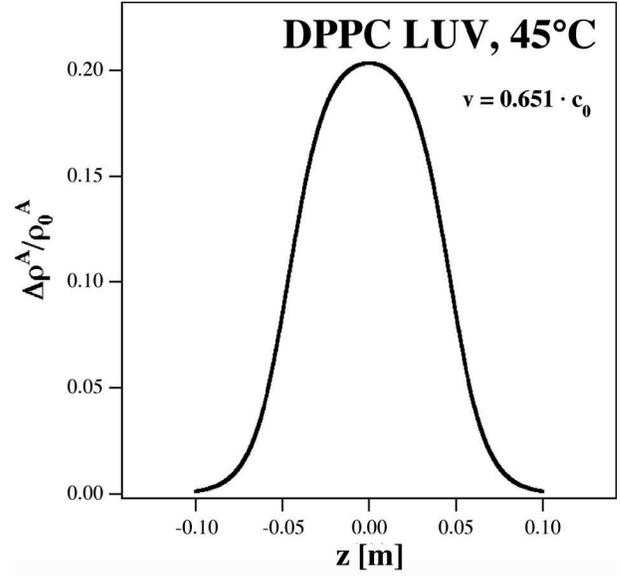}
	\parbox[c]{8cm}{ \caption{\textit{Soliton profile for a
	soliton velocity of v=0.651 c0 calculated for h=2m4/s2.  This
	soliton has a maximum amplitude of ÆrA/r0A. Its width is
	approximately 10 cm.}
	\label{fig:Figure7}}}
    \end{center}
\end{figure}
%^^^^^^^^^^^^^^^^^^^^^^^^^^^^^^^^^^^
We have shown above that the sound velocity is a function of the area
density, $\rho^A$. Introducing eq. (\ref{4.6}) into eq. (\ref{4.7}) we obtain
\begin{eqnarray}
    \frac{\partial^2}{\partial 
    t^2}\Delta\rho^A&=&\frac{\partial}{\partial 
    x}\left[\left(c_{0}^2+p(\Delta \rho^A)+q(\Delta \rho^A)^2\right)
    \frac{\partial}{\partial x}\Delta 
    \rho^A\right]\nonumber\\
    &&-h\frac{\partial^4}{\partial x^4}\Delta\rho^A
    \label{4.8}
\end{eqnarray}
and after the coordinate transformation $z=x-v\cdot t$ (introducing the
propagation velocity, $v$) we arrive at the time independent form
describing the shape of a propagating density excitation:
\begin{eqnarray}
    v^2\frac{\partial^2}{\partial 
    z^2}\Delta\rho^A&=&\frac{\partial}{\partial 
    z}\left[\left(c_{0}^2+p(\Delta \rho^A)+q(\Delta 
    \rho^A)^2\right)\frac{\partial}{\partial z}\Delta 
    \rho^A\right]\nonumber\\
    &&-h\frac{\partial^4}{\partial z^4}\Delta\rho^A
    \label{4.9}
\end{eqnarray}
This equation has a localized analytical solution \cite{Lautrup2005}:
 \begin{equation}
     \Delta 
     \rho^A(z)=\frac{p}{q}\cdot\frac{1-\left(\frac{v^2-v_{min}^2}
     {c_{0}^2-v_{min}^2}\right)}
     {1+\left(1+2\sqrt{\frac{v^2-v_{min}^2}
     {c_{0}^2-v_{min}^2}}\cosh\left(\frac{c_{0}}{h}z\sqrt{1-\frac{v^2}{c_{0}^2}}\right)\right)}
     \label{4.10}
 \end{equation}
Such localized solutions are known as solitary waves or solitons. A
typical soliton profile is shown in Fig. 7. The minimum velocity vmin
allowed by eq. (\ref{4.10}) is found to be 
\begin{equation}
    v_{min}=\sqrt{c_{0}^2-\frac{p^2}{6q}}\;.
    \label{4.11}
\end{equation}
The minimum velocity for a soliton in DPPC membranes is found to be
$v_{min}=115$ m/s, which is very close to the velocity of the action
potential found in myelinated nerves. The minimum velocity is the
velocity of the soliton when its amplitude reaches the maximum value
of
\begin{equation}
    \Delta \rho_{max}^A=\frac{|p|}{q}  ,        
    \label{4.12}
\end{equation}
corresponding to an overall density change of $\Delta
\rho_{max}^A/\rho_{0}^A=0.21$.  Solitons with larger density change do
not exist.  The total area change when going through a melting
transition is $\Delta \rho_{max}^A/\rho_{0}^A=0.246$ (for DPPC).
Thus, at maximum amplitude the soliton forces the lipid membrane by
about 85\% through the melting transition.  This will cause a
transient heat release corresponding to 85\% of the melting enthalpy
(which is on the order of 35kJ/mol or $\approx$ 13 kT per lipid).
Simultaneously, the thickness of the membrane will change by 85\% of
the thickness change in the transition from fluid to gel (7.4 \AA for
DPPC).  Since the soliton is linked to changes in lipid state the
fluorescence anisotropy will also change.  It is well known that the
anisotropy (related to the rotational mobility) is higher in the gel
phase then in the fluid phase.  Precisely these changes have all been
found in real nerves under the influence of the action potential
\cite{Tasaki1969a,Kobatake1971} (see sections \ref{sec3.1},
\ref{sec3.2} and \ref {sec3.3}).  The predicted order of magnitude of
these changes matches the data found for such nerves.

%'''''''''''''''''''''''''''''''''''''''''''''''''''''
\subsection{Electromechanical coupling}\label{sec4.4}
It seems evident that the solitons described above have many
similarities with real nerve pulses and can describe their
thermodynamic properties well. However, the action potential is known
to be a propagating voltage pulse with a net voltage change of about
100mV. In the following we will argue that this voltage change is a
consequence of the change in area density of the membrane in a manner
similar to the propagation of a piezoelectric wave. 
The membranes of biological membranes contain charged lipids.
Depending on cell and organelle the fraction of charged lipids is
between 10\% and 40\%. Some membranes are especially rich in charged
lipids, e.g. mitochondria. Typically, most of these charged lipids
are found on the inner membrane, generating an electrical field. To
make an estimate of the size of the potential change, we therefore
assume that the inner membrane of a nerve contains 40\% charged lipids
and the outer membrane contains only a very small fraction of charged
lipids (average of both leaflets 20\%). We ignore the contributions
from proteins that clearly are also present.  According to the
Gouy-Chapman theory for the potential of surfaces in electrolytes, the
potential of a charged surface at high ionic strength is given by 
\begin{equation}
    \Psi_{0}=\frac{1}{\epsilon_{0}\epsilon\kappa}\sigma\;.
    \label{4.13}
\end{equation}
This is the low potential limit of the Gouy-Chapman theory 
\cite{Trauble1976}.
The dielectric constant in vacuum is $\epsilon_{0}=$ $8.859\cdot 10^{-12}
$ $C^2/Jm$, and the relative permittivity $\epsilon=80$ for water.  Here,
$\kappa$ is the Debye constant that depends on the ionic strength.
For a monovalent salt it is given by 
\begin{equation}
    \kappa=\sqrt{\frac{2\,e^2}{\epsilon_{0}\epsilon kT}c}
    \label{4.14}
\end{equation}
where $e=1.602\cdot 10^{-19}$ C is
the elementary charge and c is the concentration of the monovalent
salt.  For c=150mM NaCl the Debye constant assumes a value
$\kappa=1.26\cdot 10^9m^{-1}$.  For a fixed number of charged lipids the charge
density, $\sigma$, is different in the fluid and in the gel phase of the
lipids because the respective lipid areas differ by
about 24\%. Therefore, one expects changes in the electrostatic
potential of the membrane during a propagating density pulse. In
piezoelectrics, voltage changes and density changes are tightly
coupled. Such coupling between lateral density and electrostatic
potential is also known as electromechanical coupling. It is also
linked to changes in capacitance. Electromechanical coupling in
membranes was first proposed by Petrov \cite{Petrov1975,Petrov1986} and has been discussed by
various authors as relevant in hair cells \cite{Petrov1994,Raphael2000}.
Here, the potential of the lipid membrane is discussed. A biological
membrane contains on average 50 weight percent of protein, which also
carry charges. The total potential of the inner and outer leaflet is
the sum of lipid and protein contributions. The contribution of the
proteins will lead to an equilibrium resting potential of the total
membrane that is different from that of the pure lipid membrane.
However, it is most likely that only the lipids undergo changes in
area during the pulse.
The potential of the inner membrane at the lipid surface under the
above conditions and the simplifying and somewhat arbitrary
assumptions regarding lipid distribution is
\begin{eqnarray}
    \Psi_{0,fluid}^{in}=-114\,mV &\qquad& 
    \Psi_{0,fluid}^{out}=0\,mV\nonumber\\
    \Psi_{0,gel}^{in}=-114\,mV &\qquad& 
    \Psi_{0,gel}^{out}=0\,mV
    \label{4.15}
\end{eqnarray}
resulting in a voltage change of $\Delta \Psi_{0}\approx 40mV$ at the
soliton peak.  That is of the same order as the voltage changes in the
action potential (which is about 100mV).  This is a very rough
estimate since the exact charge of the lipid membrane on both sides of
the membrane is not known and protein charges have not been
considered.  However, it seems as if the changes in the membrane area
during the action potential are of the right order to account for the
observed voltage changes during the action potential.  Furthermore,
membrane thickness changes during the action potential, thereby
changing the capacitance.  The assumption of a constant capacitance,
as made by Hodgkin-Huxley, therefore cannot be correct (cf.  eqs.
\ref{2.1} and \ref{2.2}).  In summary, it seems plausible that
mechanical solitons can generate voltage changes comparable to those
observed during the action potential.  The exact values remain to be
determined by experiment.

%------------------------------------------------------------------
\section{Anesthesia}
If one assumes that the soliton model for the nerve pulse is a valid
description of the nerve pulse containing its thermodynamics one
immediately arrives at a quantitative explanation for anesthesia
\cite{Heimburg2006c}.  Anesthesia as a tool for painless surgery by
use of diethyl ether was first publicly demonstrated in 1846 by
William Morton from the Massachusetts General Hospital
\cite{Campagna2003}.  This method was adopted within short time all
over the world.  Many other anesthetics had been studied in the
following decades, including both gaseous (e.g. nitrous oxide =
laughing gas) and liquid anesthetics (e.g. the alkanols from ethanol
to decanol).  A large variety of chemically distinct molecules also
cause anesthesia, e.g. barbiturates or halogenated alkanes.

%'''''''''''''''''''''''''''''''''''''''''''''''''''''
\subsection{The Meyer-Overton rule}  
About 50 years after Morton, Meyer \cite{Meyer1899} and Overton
\cite{Overton1901,Overton1991} independently found that the critical
anesthetic dose of anesthetics is linearly proportional to their
solubility in olive oil.  The critical anesthetic dose (or $ED_{50}$)
is defined as the bulk concentration of anesthetic in the air (in this
case equivalent to partial pressure) or in water at which 50\% of the
organisms studied are motionless.  Overton suggested that this finding
was related to the solubility of the molecules in the cell membrane
whose structure was not known at the time.  The Meyer-Overton rule
covers a large range of anesthetics with membrane partition
coefficients ranging over 5-6 orders of magnitude, from laughing gas
(N$_{2}$O) and the noble gas Xenon, the liquid alcohols to modern
anesthetics such as liducaine.  The partition coefficients of all
these molecules lie within error on a straight line with slope Ð1 when
plotted versus critical anesthetic dose (see Fig.  8, left).

Even after more than 160 years the effect of anesthetics on organisms
remains unexplained.  A number of functions of cells are affected by
anesthetics, including the membrane permeability, hemolysis, nerve
function and the function of ion channels and proteins totally
unrelated to anesthesia, e.g. firefly luciferase.  Since the most
obvious effect is on consciousness much of the research has focused on
the action of anesthetics on nerves.  The Hodgkin-Huxley model
\cite{Hodgkin1952} is based on the opening and closing of ion
channels, and it seems straightforward to investigate the action of
anesthetics on ion channels.  In fact, it has been observed that some
ion channel properties are influences by anesthetics.  However, this
effect is not quantitative and does not follow the Meyer-Overton rule.
Some channels are affected by some anesthetics but not by others.  As
an example, voltage gated sodium and potassium channels are slightly
inhibited by halogenated alkanes and ethers but not by Xenon and
nitrous oxide, although all these anesthetics follow the Meyer-Overton
rule in causing anesthesia \cite{Campagna2003}.  It has to be
concluded that protein pictures of anesthesia are not yet
satisfactory.
 
%vvvvvvvvvvvvvvvvvvvvvvvvvvvvvvvvvvv
\begin{figure*}[htb!]
    \begin{center}
	\includegraphics[width=12cm]{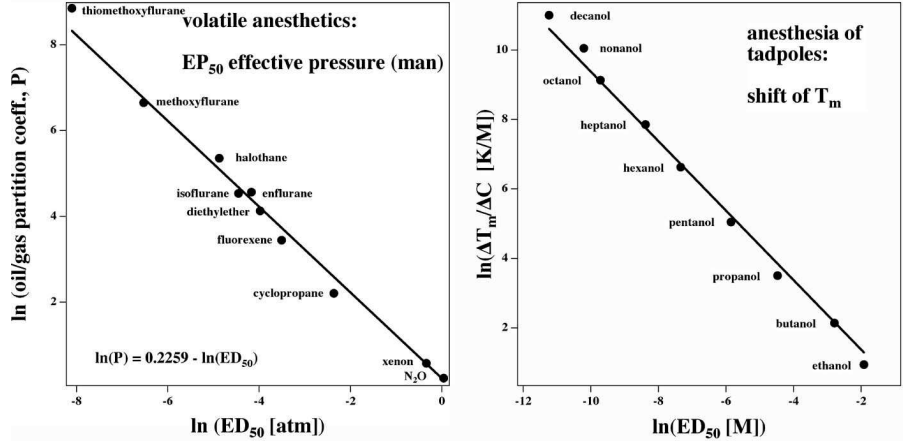}
	\parbox[c]{14cm}{ \caption{\textit{Left: The Meyer-Overton
	rule for volatile anesthetics showing the linear dependence of
	the oil/gas partition coefficient and the critical anesthetic
	dose for man.  The solid line represents a straight line with
	slope Ð1.  Data adapted from ref.  27.  Right: Lowering of the
	melting transition for a series of alkanols as a function of
	the critical anesthetic dose for tadpoles.  The solid line
	displays a slope of Ð1.  Adapted from ref.  46.}
	\label{fig:Figure8}}}
    \end{center}
\end{figure*}
%^^^^^^^^^^^^^^^^^^^^^^^^^^^^^^^^^^^
The Meyer-Overton rule suggests that the effect of anesthetics is
independent of the chemical nature of the molecule. Since the noble
gas Xenon lies on the same straight line as halothane or the liquid
anesthetics, one can essentially rule out specific binding effects,
which are the basis of the protein models (see also discussion).

%'''''''''''''''''''''''''''''''''''''''''''''''''''''
\subsection{Melting point depression}
It is known that anesthetics have a pronounced effect on lipid melting
transitions.  Typically, with addition of anesthetics to the bilayers,
transitions shift to lower temperatures in a linear relation with the
anesthetic concentration.  Heimburg and Jackson \cite{Heimburg2006c}
have shown that this effect can be described by accurately by the
well-known phenomenon of freezing point depression.  If one assumes
that anesthetics molecules are readily soluble in fluid lipid
membranes and insoluble in the gel membrane, one arrives at following
law for the freezing point depression
\begin{equation}
    \Delta T_{m}=-\frac{RT_{m}^2}{\Delta H}x_{A}  \;,
    \label{5.1}
\end{equation}
where $x_{A}$ is the molar fraction of anesthetics in the fluid lipid
membrane, $T_{m}$ is the melting point of the lipid membrane and ÆH is the
melting enthalpy. The derivation of this equation can be found in any
physical chemistry textbook. The membrane concentration of
anesthetics at the critical dose, $x_{A}$, is related to the partition
coefficient via
\begin{equation}
    x_{A}=P\cdot(ED_{50})\cdot V_{l}
    \label{5.2}
\end{equation}
where $P$ is the partition coefficient between membrane and water,
$ED_{50}$ is the critical anesthetic concentration, and $V_{l}$ is the
molar volume of the lipids (about 0.75 l/mol).  The two equations
above describe the behavior of many anesthetics.  D. Kharakoz
\cite{Kharakoz2001} has collected data for various anesthetics, some
of which are displayed in Fig.  8 (right).  Shown is the concentration
dependence of the melting point as a function of the critical
anesthetic dose for tadpoles.  The melting point depression for all
anesthetics (shown here are alkanols) lie on a straight line when
plotted versus the critical anesthetic dose.  The slope of the curve
indicates that the shift of the transition temperature at critical
anesthetic dose is $\Delta T_{m}=-0.6$ K for all anesthetics that
follow the Meyer-Overton rule, independent of the chemical nature of
the drug \cite{Heimburg2006c,Kharakoz2001}.  The Meyer-Overton rule
therefore can be reformulated as: The anesthetic potency of
anesthetics is proportional to their ability to lower the melting
point of lipid membranes.  It is clear that within the soliton model
for nerve pulses the melting points play an essential role.  The
assumption in the following is that the lipid melting point plays an
important role in the control of biological membranes.

%'''''''''''''''''''''''''''''''''''''''''''''''''''''
\subsection{Pressure reversal}
If one assumes that the lipid melting point is important for
biological function and that the effect of anesthetics is related to
their effect on melting points, it is interesting to compare this to
other physical properties that also influence melting points, most
notably the influence of pressure. It has long been known that
pressure influences the melting points of membranes by shifting them
to higher temperatures. The pressure dependence of such transitions
is described by \cite{Ebel2001} 
\begin{equation}
    \Delta T_{m}=\gamma_{V}\Delta p\,T_{m}
    \label{5.3}
\end{equation}
where $\gamma_{V}=7.8\cdot 10^{-10}m^2/N$ is a constant that is roughly the
same for all lipids, lipid mixtures and biological membranes
\cite{Heimburg1998,Ebel2001}.  This equation indicates that a lipid
membrane with a $T_{m}=314$ K (dipalmitoyl phosphatidylcholine) shifts
its transition by 1 K to higher temperatures upon application of 40.8
bar hydrostatic pressure.  This indicates that a pressure of 24.5 bar
should be sufficient to reverse the effect of anesthesia (which
corresponds to a shift by 0.6 K to lower temperatures).  Pressure
reversal of anesthesia has indeed been found, first by Johnson et al.
\cite{Johnson1950}.  If tadpoles are anesthetized at 3 times the
critical anesthetic dose of ethanol, they wake up upon application of
150 bars of hydrostatic pressure.  The pressure reversal of anesthesia
is well documented in the literature.

%'''''''''''''''''''''''''''''''''''''''''''''''''''''
\subsection{Free energy of the membrane}
The free energy difference between gel and fluid phase is the free
energy that must be provided to shift the lipid membrane through its
phase transition. It is given by
\begin{equation}
    \Delta G=\Delta H-T\Delta S=\Delta H\cdot\left(\frac{T_{m}-T}{T_{m}}\right)   ,
    \label{5.4}
\end{equation}
making use of the identity $\Delta S=\Delta H/T_{m}$.  This equation
indicates that the free energy difference between the two phases is
linearly dependent on the difference of the experimental temperature,
$T$, and the melting temperature, $T_{m}$.  Now, we have shown in the
previous section that $T_{m}$ is influenced by both anesthetics
concentration and by pressure.  The melting temperature $T_{m}$ is changed
by anesthetics and pressure in the following manner 
\begin{equation}
    T_{m}=T_{m,0}-\frac{RT_{m,0}^2}{\Delta H}x_{A}+\gamma_{V}\Delta 
    p T_{m,0}
    \label{5.5}
\end{equation}
where $T_{m,0}$
is the transition temperature at atmospheric pressure and in the
absence of anesthetics.  We finally obtain
\begin{equation}
    \Delta G(x_{A},\Delta p)\approx \Delta 
    H\left(\frac{T_{m,0}-T}{T_{m,0}}-\frac{RT}{\Delta 
    H}x_{A}+\gamma_{V}\Delta p\frac{T}{T_{m,0}}\right)\,.
    \label{5.6}
\end{equation}
If the melting
transition of the lipid membrane is to play a relevant role for
biological function, it follows that biological function should be the
same when $\Delta G$ is the same.  Therefore, the condition for pressure
reversal of anesthesia is or
\begin{eqnarray}
    \Delta p&\approx& \frac{1}{\gamma_{V}}\frac{RT_{m,0}}{\Delta H}x_{A}
    \nonumber\\
    \Delta p_{ED_{50}}&\approx& 
    \frac{1}{\gamma_{V}}\frac{RT_{m,0}}{\Delta H}P(ED_{50})V_{l}\,. 
    \label{5.7}
\end{eqnarray} 
The numbers obtained from this equation are of an order very similar
to that obtained in experiments.  Data from octanol and DPPC membranes
as well as the equations above suggest that a pressure of 24.5 bar
reverses anesthesia \cite{Heimburg2006c}.  The data from Johnson on
tadpoles in an ethanol solution corresponding to three times the
anesthetic dose was reversed by 150 bars of pressure
\cite{Johnson1950}.  Our calculation yields 73.5 bars, assuming a
membrane partition coefficient for ethanol of 0.6 (which is subject to
an error of the order of a factor 2).

%------------------------------------------------------------------
\section{Discussion}
We have suggested here that the Hodgkin-Huxley model
\cite{Hodgkin1952} for the action potential does not provide a
satisfactory description of nervous impulse because it does not
include the mechanical and optical changes associated with the action
potential.  Further, it is clearly inconsistent with the thermal
response.  The initial heat release and subsequent re-absorption
studied by a number of authors
\cite{Abbott1958,Howarth1968,Ritchie1985,Tasaki1989,Tasaki1992a} points
rather to a reversible physical phenomenon that conserves entropy.  In
contrast, the Hodgkin-Huxley model is based on the flux of currents
through resistors that should heat the membrane independent of the
nature of the current and its direction.  Therefore, we have proposed
an alternate model based on the known mechanical and thermal features
of artificial and biological membranes.  It was shown that under
physiological conditions stable mechanical solitons could propagate
and display reversible heat release, changes in membrane thickness,
changes in membrane order and reversible membrane potential changes.
All these changes have been observed in experiments.  In particular
the reversible heat release and the overall conservation of entropy is
a feature typical of sound propagation.  It should be noted that we
use the term Ôsound propagationÕ in a general sense that includes all
changes of the thermodynamic variables that accompany a mechanical
compression according to MaxwellÕs relations.  In such a description,
the simultaneous occurrence of density changes, voltage changes, and
heat release is a surprise but rather a necessary consequence of
thermodynamics.  The Hodgkin-Huxley model seems to be in agreement
with fluxes through ion channel proteins.  However, the currents
through such channels fall short of presenting an explanation in the
sense of a physical theory based on first principles.  The
conductances of the channels contain many parameters that cannot be
justified theoretically.  Therefore, their seemingly simple
description relies on objects that contain all the unexplained
features in the form of parameters.  For this reason Hodgkin and
Huxley originally recommended treating their model with care.  They
state in their seminal paper from 1952
\cite{Hodgkin1952}:\textit{''The agreement must not be taken as
evidence that our equations are anything more than an empirical
description of the time-course of the changes in permeability to
sodium and potassium.  An equally satisfactory description of the
voltage clamp data could no doubt have been achieved with equations of
very different form, which would probably have been equally successful
in predicting the electrical behavior of the membrane.  \ldots the
success of the equations is no evidence in favour of the mechanism of
permeability change that we tentatively had in mind when formulating
them.''} In this paper we have, in fact, shown that many changes can
be explained by totally different physical mechanisms that result in
similar equations for the pulse propagation.  Hodgkin was clearly
aware of the problems generated by the finding of a reversible heat
releases during the action potential.  He wrote in his textbook `The
conduction of the nervous impulse' \cite{Hodgkin1964}: \textit{``In
thinking about the physical basis of the action potential perhaps the
most important thing to do at the present moment is to consider
whether there are any unexplained observations which have been
neglected in an attempt to make the experiments fit into a tidy
pattern.  \ldots perhaps the most puzzling observation is one made by
A.V. Hill and his collaborators Abbott and Howarth
(1958).\cite{Abbott1958} \ldots Hill and his colleagues found that it
(the heat release) was diphasic and that an initial phase of heat
liberation was followed by one of heat absorption.  É a net cooling on
open-circuit was totally unexpected and has so far received no
satisfactory explanation.''} Howarth et al.  \cite{Howarth1968}
concluded from their finding of heat release and subsequent heat
uptake:\textit{''It seems probable that the greater part of the
initial heat results from changes in the entropy of the nerve membane
when it is depolarized and repolarized.''} Reversible entropy changes,
however, are not a feature of textbook pictures of nerve pulses.
Here, we have followed HodgkinÕs suggestion and searched for ways to
explain the reversible heat.  Slightly below physiological
temperatures, there exist chain-melting transitions of the membrane.
It is interesting to note that these transitions occur at much lower
temperatures in the absence of the proteins.  For instance, the
melting point of \textit{E.coli} lipid extracts is about 20K lower
than that of the native membrane in the presence of all their lipids.
Therefore, the presence of proteins seems to play an essential role in
fine-tuning the thermodynamics of biological membranes.  Besides their
role as catalysts, proteins also possess chemical potentials that are
thermodynamics variables.  They contribute to the behavior of
membranes in a manner similar to temperature, pressure, pH and other
variables.  The presence of cooperative lipid transitions forms the
basis for the possibility of density pulses that propagate along the
nerve axon.  One short-coming of our model is that it does not yet
include a frictional term even though one may expect that, due to the
flux of lipids and changes in diameter of the nerve, a proper
hydrodynamic treatment should yield in a dampening of the pulse.  This
problem remains unanswered in the context of our model, mainly due to
the lack of detailed data on the dimensional changes in nerves.
However, experiments show that such density pulses propagate
\cite{Iwasa1980a,Iwasa1980b} in real nerves, and the near-complete
reversal of the heat \cite{Abbott1958,Howarth1968,Ritchie1985}
suggests that friction is small.  Within the soliton model proteins do
not play a role as channels or as active components.  Rather, they
tune the thermodynamics of the membrane.  An important question is how
such a mechanical soliton can be generated in a membrane.  Since the
soliton pushes the membrane through its chain melting transition,
everything that moves membranes through transitions should be able to
generate a pulse.  All physical changes that push the transition away
from physiological conditions should inhibit pulses.  As an example,
local cooling of a nerve has been shown to induce nerve firing,
whereas temperature increase inhibits pulse conduction
\cite{Kobatake1971}.  Due to the electromechanical coupling described
in section \ref{sec4.4}, changes in trans-membrane voltage are also
potentially able to generate pulses.  Further, a local decrease of pH,
increase in pressure or increase in calcium concentration all have the
potential to trigger pulses because all of these changes increase the
phase transitions of biomembranes.  Most interestingly, anesthetics
inhibit pulse generation due to their property of lowering phase
transitions.  Since ion channels do not play an active role in our
description of nerve pulses, it is obvious that the action of
anesthetics requires a different explanation than their action on ion
channels.  The famous 100-year old Meyer-Overton correlation
\cite{Meyer1899,Overton1901,Overton1991,Campagna2003} states that the
action of anesthetics is, within error, strictly proportional to their
solubility in lipid membranes.  This law is valid over 6 orders of
magnitude in the membrane/air and membrane/water partition
coefficient.  This law remains an elegant and valid means to determine
the effectiveness of an anesthetic \cite{Campagna2003}.  It basically excludes the
notion that the action of anesthetics can be linked to specific
binding of the drug to a receptor.  The argument is simple: The
binding of two molecules is described by the free energy, which is a
function of state.  If the action of anesthetics is exactly
proportional to the concentration of drugs in the membrane independent
on chemical nature of the drug as follows from the Meyer-Overton
correlation, the binding constant of all anesthetics to receptors must
be identical, including that of the noble gas Xenon.  Since noble
gases cannot bind specifically, the same must be concluded for all
other anesthetics that follow the Meyer-Overton rule.  The
experimental finding is that halogenated alkanols act very differently
on ion channels than Xenon or nitrous oxide.  Thus, protein models are
clearly not consistent with the well-documented Meyer-Overton
correlation.  Although protein models are currently quite popular,
they cannot fulfill the basic thermodynamic requirements for
anesthetics that follow the Meyer-Overton correlation.  Due to the
above argument it is unlikely that the action on ion channels is
related to anesthesia.  Here, we have outlined the thermodynamic
theory of how anesthetics influence the phase behavior of lipid
membranes via a well-known unspecific phenomenon known as
freezing-point depression.  It states the lowering of the melting
point is proportional to the membrane concentration of the anesthetic
drug.  Thereby, we attribute a physical meaning to the Meyer-Overton
rule that was not provided by Overton himself.  By this mechanism
anesthetics alter the features of propagating solitons in a
quantitative manner.  More specifically, they alter the amount of free
energy that has to be provided to generate a pulse.  We found that it
is linearly dependent on the distance between physiological
temperature and the transition in the nerve membrane.  This approach
admits the possibility of finding strict thermodynamics relations
between various thermodynamics variables, including the pressure
reversal of anesthesia that can be calculated in quantitative terms 
\cite{Johnson1950}.
It seems unlikely that all of these quantitative correlations can be
found experimentally without thermodynamics being an essential player
in the description of both the action potential and the action of
anesthesia.  Indeed, simple thermodynamics seems to contain a complete
description of such phenomena.  \\

\textbf{Acknowledgments:} We thank Dr. E. Neher (Max Planck Institute,
G\"{o}ttingen) for discussing the problem of friction of the solitons,
and Dr. I. Tasaki (NIH, Bethesda) and Dr. K. Kaufmann (G\"{o}ttingen) for
helpful discussion on the mechanical nature of nerve pulses. Dr. B.
Lautrup (NBI, Copenhagen) contributed the analytical solution of the
propagating soliton and many hours of discussion.

\footnotesize
%\bibliography{literature201006bibdesk}
\bibliographystyle{biophysj2005}

\end{document}